\documentclass[twocolumn,aps,showkeys]{revtex4-2}

\usepackage{color}
\usepackage{xcolor}
\usepackage{multirow}
\usepackage{amsmath}
\usepackage{esvect}

\usepackage{import}
\usepackage{graphicx}
\usepackage{amsmath}
\usepackage{array}
\usepackage{bm}

\usepackage{amsfonts}    
\usepackage{amssymb}
\usepackage{latexsym}
\usepackage[toc,page]{appendix}
\usepackage{graphicx}
\usepackage[english]{babel}

\usepackage[colorlinks=true,urlcolor=blue,citecolor=blue,linkcolor=blue]{hyperref}
\newcommand{\len}{\Delta}
\newcommand{\ind}{\alpha}
\newcommand{\Nint}{D}

\newcommand{\diff}{\mathop{}\!\mathrm{d}}

\begin{document} 
\widetext

\title{A numerical study of the Bose-Einstein condensates in a double-well trap using finite differences}
\author{D~.~J.~ Nader}
 \email{daniel.juliannader@upol.cz} 
\affiliation{Department of Optics, Faculty of Science, Palack\'y University, Olomouc 77146, Czech Republic}
\author{E. Serrano-Ensástiga}
\affiliation{Institut de Physique Nucléaire, Atomique et de Spectroscopie, CESAM, University of Liège
\\
B-4000 Liège, Belgium}

\date{\today}
%
%
%
\begin{abstract}
Bose-Einstein condensates in a double-well potential contain the essential ingredients to study many-body systems within a rich classical phase-space that includes an unstable point and a separatrix. Employing a selfconsistent finite difference method, we study some of their quantum properties and their dependency on the strength of the boson-boson interaction. We observe a deviation in the critical parameters associated with a behavior change in both the energy distribution and the eigenstates of the system. We also examine the trends of the nonclassicality via the Wigner function, the tunneling transmission coefficient, and the nonorthogonality of eigenstates associated with the nonlinearity aspects of the Gross-Pitaevskii equation. 
\end{abstract}


\maketitle

\section{Introduction}
The double-well potential is one of the favorite confinement traps to explore the peculiar properties of the spectrum and dynamics of a quantum system~\cite{vybornyi2014tunnel,Prado2023,Schmelcher2006}. The symmetry breaking over the origin caused by the extra barrier in the double-well potential~\cite{Joger,DELABAERE1997180,meier,Dong2019} produces two global minima which served as a model to test tunneling effects and quantum-classical correspondence~\cite{Plata_1992,Novaes_2003,Murmann2015,Weiss,Ratelli,Lu}. Additionally, the presence of a fixed unstable point (local maximum) and a separatrix in the classical phase-space, which separates two different classes of solutions in the classical system, have consequences in its corresponding quantum system such as a discontinuity in the energy distribution and a different shape of the wavefunctions of the eigenstates over a critical energy $E_c$~\cite{NADER2023129014}. Moreover, the dynamics of quantum states with probability densities localized near to the separatrix are more susceptible to tunneling transitions and an exponential growth in the out-of-time-ordered correlators~\cite{hasegawa2013gaussian,Cary,chavez2023spectral}. All the previous phenomena described above are associated to an Excited State Quantum Phase Transition (ESQPT) in the discrete spectrum \cite{cejnar2021excited}. The theoretical predictions of quantum systems confined in double-well potentials have been tested and used in experiments with superconducting circuits~\cite{Harris2008}, periodically driven quantum systems~\cite{venkatraman2023driven,Kiering,Dykman2007} and cold atom systems~\cite{PhysRevA.73.033619,hofferberth2006radiofrequency}. 

The Bose-Einstein condensates (BEC) offer unprecedented opportunities to study the underlying physics mentioned above due to the astonishing precision and control levels achievable in experiments~\cite{pethick2008bose,lewenstein2012ultracold}. One of the main examples is the Josephson junction implemented by two linked Bose-Einstein condensates in a double-well potential, where macroscopic tunneling oscillations are observed~\cite{PhysRevLett.95.010402,schumm2005matter}. Analog BECs systems in double-well potentials have been also explored, such as spinor BECs~\cite{PhysRevLett.83.661,Gomez2007,Eto}, or toroidal BECs, also called atom SQUIDS~\cite{PhysRevLett.111.205301}. On the theoretical side, the spectra, dynamics and tunneling phenomena of BECs confined in double-well potentials have been also studied~\cite{Alon2009,kenkre2022bose,andersen2022quantum,Schmelcher2010,Hu_2019,shch2008,Prates2022,Bello2017,PhysRevLett.94.020404}, including double wells generated by Gaussian functions~\cite{Schmelcher2008} or by a quartic polynomial~\cite{Jaaskel}. Since the many-body problem is quite challenging in the general case, simplifications are made such as the use of 1-Dimensional (1D) condensates, which have been realized experimentally~\cite{Science.1DGP,PhysRevLett.96.130403}, and the mean-field approximation that derives the well-known Gross-Pitaevskii (GP) equation~\cite{Gross,Pitaevskii,pethick2008bose}. Another well-known theoretical approach is the finite multimode model, where it is considered a truncated Hilbert space for the single quantum states of the cold atoms~\cite{PhysRevA.61.031601,pethick2008bose,ananikian2006gross}. On the numerical side, the Finite Difference method for the study of 1D-BEC has no developed~\cite{Trofimov}, and fully studied in a recent review~\cite{halpern2022visualizing}. In particular, it shows some advantages with respect to other modern numerical methods~\cite{Trofimov}.

In this work, we exploit the Finite Difference method to obtain the stationary solutions and their respective eigenvalues of the 1D-BEC in the mean-field approximation. Following the compilation in Sec.~\ref{Sec2.theory} of the theoretical elements needed in this work, we expose the numerical method in Sec.~\ref{Sec3.Method}. We then proceed to study several physical properties associated to a quantum many-body system in a double-well potential in Sec.~\ref{Sec4.Results}. Specifically, we study the relation between the strength of the boson-boson interaction with the critical energy $E_c$ of the BEC, which in the non-interacting case coincides with the local maximum of the double well. We also discuss some consequences on the nonclassicality, the tunnelling transmission coefficient in the static perspective, and the nonorthogonality of the eigenstates. Section~\ref{Sec5.Conclusions} summarizes our findings and offers some thoughts on future work.
\section{Theory}
\label{Sec2.theory}
Let us considered a 1D-BEC gas of $N$ atoms, with mass $m$, and at low energies such that the predominant boson-boson interactions are produced by point-contact collisions~\cite{pethick2008bose}. We also adopt the mean-field approximation that assumes the fully condensed wavefunction $\Phi$ as the product of normalized single-particle wavefunctions 
\begin{equation}
    \Phi (\bm{r}_1 , \dots , \bm{r}_N) = \prod_{i=1}^{N} \Psi(\bm{r}_i) \, , 
\end{equation}
with 
\begin{equation}
   \int |\Psi(\bm{r})|^2 \diff \bm{r}= 1 \, .
\end{equation}
The GP equation of the 1D-BEC reads \cite{Gross,Pitaevskii,pethick2008bose}
\begin{equation}
 \label{GP}
 \left(-\frac{\hbar^2}{2m}\frac{d^2}{dx^2} + V(x) + \beta |\Psi_n(x)|^2\right)\Psi_n (x)=\mu_n \Psi_n(x) \, ,
\end{equation}
where $m$ is the mass of the identical bosons, $\mu_n= \mu_n'/N$ the chemical potential per particle and $\Psi_n(x)$ the wave function of the $n$th-eigenstate of the condensate. The coupling factor $\beta$ of the nonlinear term in the GP equation is proportional to the $s$-wave scattering length of the boson-boson interaction $a_s$, $\beta= 4\pi N \hbar^2 a_s / m$. The trap considered in this work is the quartic potential in the form of a symmetric double well (Mexican hat)
\begin{equation}
\label{potencial}
V(x)=-ax^2 + b x^4
\end{equation}
where $a$ is a positive parameter. 
Since the potential is symmetric with respect to the transformations $(x\to-x)$, the 
eigenstates of the Schrodinger equation ($\beta=0$) with eigenvalues below the local maximum  of the double well are organized in quasidegenerated pairs and the energy gap is proportional to the square root of the transmission coefficient \cite{NADER2023129014}. 
The parameter $b$ of the double-well potential (\ref{potencial}) can be absorbed in the other variables by the Szymanzik rescaling~\cite{Turbiner:2009ns}
\begin{equation}
\left( \tilde{x} , \tilde{a} \, , \tilde{b} \, , \tilde{\beta} \, , \tilde{\mu} \right) = \left( b^{1/6} x \, , \frac{a}{b^{2/3}} \, , 1 \, , \frac{\beta}{b^{1/3}} \, , \frac{\mu}{b^{1/3}}  \right)
\, , 
\end{equation}
and
\begin{equation}
    \tilde{\Psi}(\tilde{x}) = \Psi(x)  \, .
\end{equation}
In the following, we will work with the scaled variables and suppress the tilde symbol in each term. Once the eigenfunctions $\Psi_n$ are known, the energy of the condensate per particle is given by the expectation value \cite{pethick2008bose,dion2007ground}
\begin{equation}
\label{energy}
E_n=\int\left( \frac{1}{2m} \left| \frac{\diff \Psi_n(x)}{\diff x} \right|^2 + V(x)|\Psi_n(x)|^2 + \frac{\beta}{2}|\Psi_n(x)|^4\right) \diff x\,.
\end{equation}

As a measure of nonclassicality, one can use the volume of the integrated negative part of the Wigner function $W_{\Psi_n}(x,p)$~\cite{AnatoleKenfack_2004}
\begin{equation}
\label{Eqn.nonclass}
  \delta(\Psi_n)=\int\int |W_{\Psi_n}(x,p)| \diff x \diff p -1\,,  
\end{equation}
with
\cite{Weinbub2018}
\begin{equation}
\label{Wigner}
W_{\Psi_n}(x,p)=\int \Psi^*_n(x-y)\Psi_n(x+y) e^{\frac{2ipy}{\hbar}}  \diff y \,.
\end{equation}
This phase-space representation of the BEC in a double well by the Wigner function is complementary to that given by the Husimi function \cite{Mahmud} which admits only positive values.

Lastly, following the WKB approximation for BEC condensates with small $\beta$~\cite{Lindberg_2023,deCarvalho}, the transmission coefficient from one well to the other reads
\begin{equation}
\label{Twkb}
\begin{aligned}
T_n&=\exp \left(-\frac{2}{\hbar}\gamma_n \right), 
\\ 
\gamma_n&=\int_{x_1}^{x_2}\sqrt{2m \left( V(x)+\beta|\Psi_n(x)|^2-\mu_n \right)} \diff x \,,    
\end{aligned}    
\end{equation}
where $x_{1,2}$ are the classical turning points in the effective potential of the GP equation~\eqref{GP}. Throughout the rest of the paper, we consider $m=\hbar= 1$.
\section{The Method}
\label{Sec3.Method}
In this work, we employ the  method of Finite-Difference discretization of the Schr\"odinger equation \cite{halpern2022visualizing} and a selfconsistent algorithm
to incorporate the nonlinear term \cite{Trofimov}. In the Finite-Difference method, the configuration space $x\in[-L,L]$ is divided in $\Nint$ subintervals of length $\len =2L/\Nint$
such that $x_{\ind}=-L+ \ind \len$, where $\ind =0,1,..,\Nint$. We take a sufficiently large $\Nint$ such that the first derivative of the wavefunction $\frac{\diff \Psi_n}{\diff x}$ at each $x_{\ind}$ is well approximated by 
\begin{equation}
\label{derivada}
\frac{\diff \Psi_n(x_{\ind})}{\diff x}\approx \frac{\Psi_n(x_{\ind+1})-\Psi_n(x_{\ind})}{\Delta} \, .
\end{equation}
Repeating this argument, we find the formula to calculate the second derivative appearing in the kinetic energy, 
\begin{equation}
\label{2derivada}
\frac{\diff ^2\Psi_n(x_{\ind})}{\diff x^2}\approx \frac{\Psi_n(x_{\ind+1})-2\Psi(x_{\ind})+\Psi_n(x_{\ind-1})}{\Delta^2}\,.
\end{equation}
By using the approximation above, the GP equation~\eqref{GP} takes the form of an eigenvalue problem 
\begin{equation}
\label{eig.ite}
A_n^{(k)}\Psi^{(k)}_n=\mu_n^{(k)}\Psi^{(k)}_n  \, ,
\end{equation}
where $A_n^{(k)}=-\frac{\hbar^2}{2m}A_1+A_2 + \beta A_{3,n}^{(k)}$ is a $(D+1)\times (D+1)$ matrix composed by the tridiagonal matrix
$$
A_1= \begin{pmatrix}
-2 & 1 &  & & \\
1 & -2 & 1 & & \\
  & \ddots & \ddots &\ddots & \\
  &        &      1 & -2   & 1 \\
  &        &        & 1    & -2 \\
\end{pmatrix} \, ,
$$
corresponding to the kinetic energy, and the diagonal matrices associated to the potential trap and the boson-boson interactions, respectively, with entries
\begin{equation*}
\left( A_{2} \right)_{rs} = V(x_{r-1}) \delta_{rs} \, , \quad 
\left( A_{3,n}^{(k)} \right)_{rs} = |\Psi^{(k-1)}_n(x_{r-1})|^2 \delta_{rs}  
\, . 
\end{equation*}
The superindex $(k)$ indicates the corresponding iteration step of each element. In particular, the matrix $A_{3,n}^{(k)}$ is built with the $n$th-eigenfunction of the previous iteration, with initial eigenvectors assumed by normalized constant functions $\Psi_n^{(0)}(x_{\ind})= \sqrt{N/2L}$. The Eq.~\eqref{eig.ite} is solved self-consistently until reaching convergence of both the eigenvalues and eigenfunctions up to the orders
\begin{equation}
\label{eq.conver}
    \left| \mu_n^{(k)} - \mu_n^{(k-1)} \right| < 10^{-9} \, , \quad 
    1- | \langle \Psi^{(k)} | \Psi^{(k-1)} \rangle| < 10^{-4} \, .
\end{equation}
The value of $L$ is considered large enough that the solution fulfills $|\Psi(\pm L)| \approx 10^{-3}$ and, consequently, its probability density outside the interval $[-L,L]$ is negligible.  Once we have calculated the respective discretized eigenfunction, any integral can be calculated approximately by 
$$\int  f(x) \diff x \approx \len \sum_{\ind=1}^{\Nint}f(x_{\ind})\,.$$
\begin{figure}[t!]
    \begin{tabular}{r}
    \includegraphics[scale=0.6]{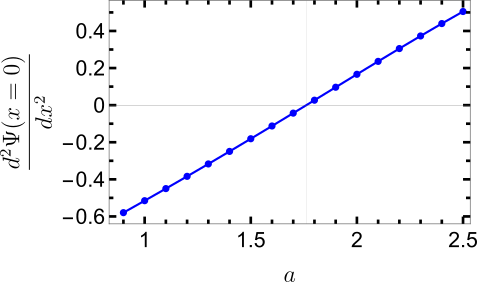} 
\\
\includegraphics[scale=0.56]{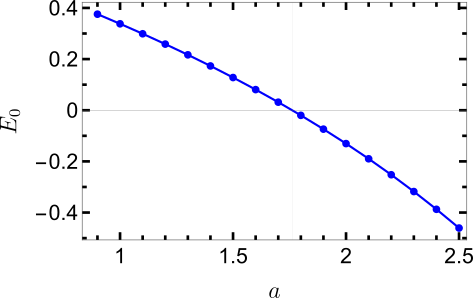} 
    \end{tabular}
    \caption{\label{Fig1}
    Energy (left) and second derivative of the ground state evaluated at the center of coordinates (right) as a function of the parameter $a$. The parameter $\beta$ is set to zero to coincide with the Schr\"odinger equation with the quartic double-well potential. The vertical gray line indicates the transition at the critical parameter, both at $a_c=1.7616$, associated with the change of sign of the respective quantity. 
    }
\end{figure}
\section{Results}
\label{Sec4.Results}
\subsection{Critical Parameter}
\begin{figure*}[t!]
    \begin{tabular}{cccc}
     \multicolumn{2}{c}{\includegraphics[width=5cm, height=4cm,angle=0]{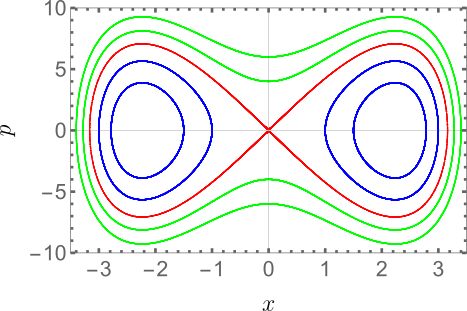} }
   &  \multicolumn{2}{c}{\includegraphics[width=6cm, height=3.6cm,angle=0]{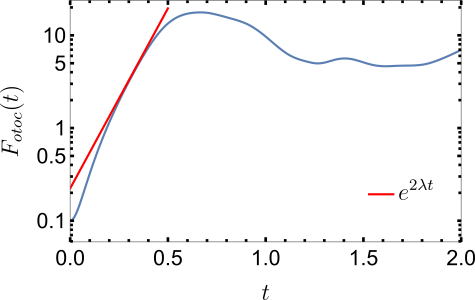}} \\
 $t=0$ &  $t=0.1$ &   $t=0.2$ &  $t=0.3$ \\
  \includegraphics[width=3cm, height=3cm,angle=0]{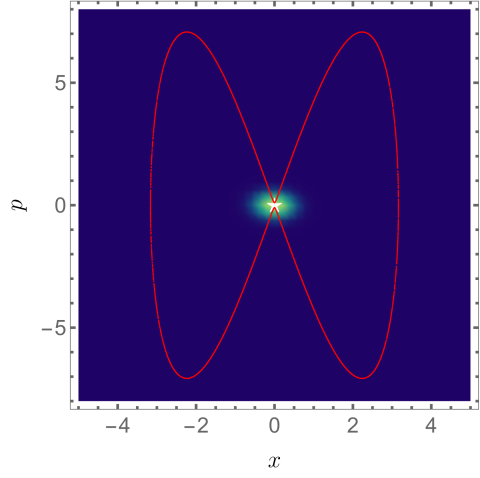} 
   &  \includegraphics[width=3cm, height=3cm,angle=0]{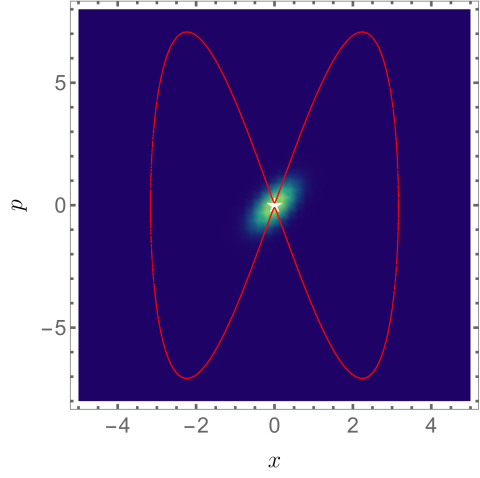}  
   & \includegraphics[width=3cm, height=3cm,angle=0]{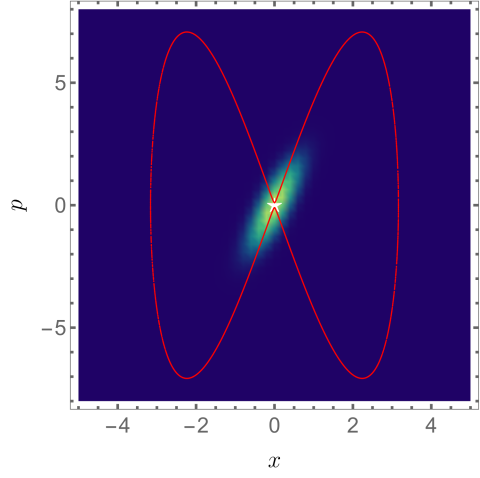}
   & \includegraphics[width=3cm, height=3cm,angle=0]{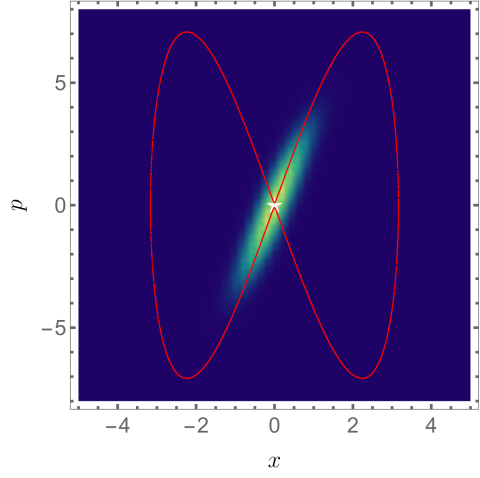}
   \\ 
    \end{tabular}
    \caption{\label{Fig2}
    (Top-left plot) Classical trajectories of the double well $(a \, ,\beta)= (10,0)$. The solid red line indicates the separatrix at the critical energy $E_c=0$, the solid blue lines are trajectories with energies below the critical and the green lines corresponds to energies above the critical. (Top-right plot) Fidelity Out-Of-Order Correlator $F_{otoc}(t)$ in Logarithmic scaled of an initial coherent state $|\alpha(x_0,p_0)\rangle$ lying along the separatrix. The slope of the exponential growth can be compared against the positive Lyapunov exponent $\lambda=\sqrt{2a}$ of the fixed point $(x_0,p_0) = (0,0)$. (Bottom plot) Time evolution of an initial coherent state initially centered at the origin.
    }
\end{figure*}

\begin{figure}
\centerline{\includegraphics[width=0.45\textwidth]{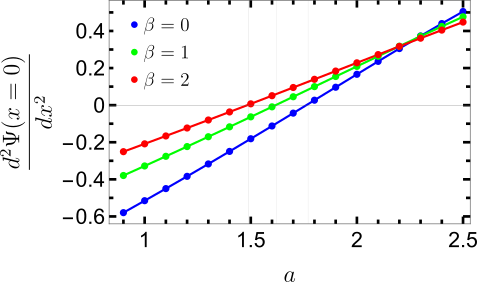}}
\caption{\label{Fig3} Second derivative of the ground-state wavefunction evaluated at the origin as a function of the parameter $a$ for different values of $\beta$.}
\end{figure}

\begin{figure}
\includegraphics[width=5cm, height=3.5cm,angle=0]{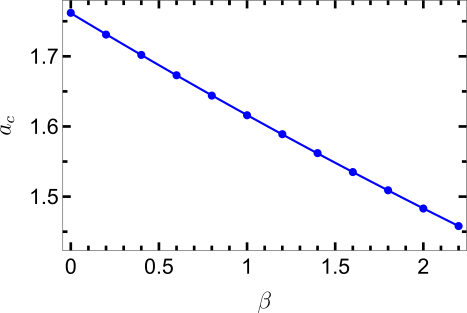}
\\
  \includegraphics[width=5cm, height=3.5cm,angle=0]{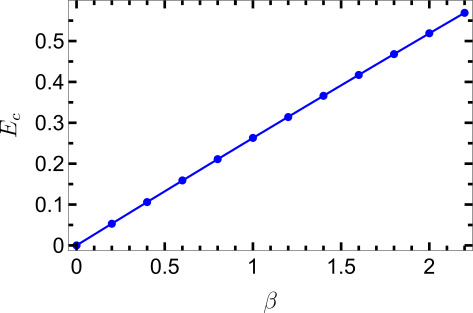}
\caption{
\label{Fig4}
    Critical parameter $a_c$ (left) and energy $E_c$ (right) as a function of the nonlinearity parameter 
    $\beta$. The dashed lines are the quadratic fits $a_c(\beta)\approx 1.7616 - 0.1513 \beta + 0.0061 \beta^2$ and $E_c(\beta)\approx 0.2662 \beta + 0.0034 \beta^2$, respectively.
    }
\end{figure}

\begin{figure*}
    \begin{tabular}{cc}
    \multicolumn{2}{c}{$a=2$}\\
    \includegraphics[width=7cm, height=5cm,angle=0]{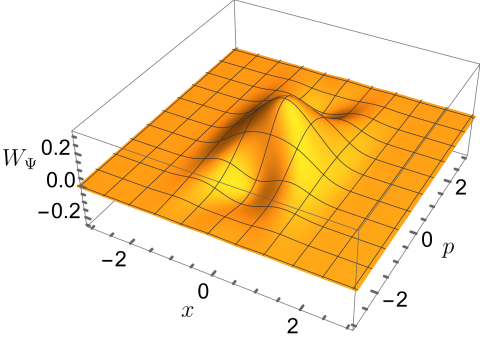} 
   &  \includegraphics[width=6cm, height=4.5cm,angle=0]{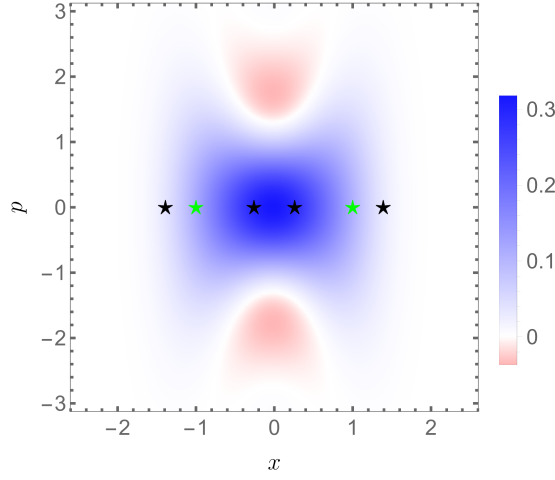} 
   \\ 
   \multicolumn{2}{c}{$a=5$}\\
    \includegraphics[width=7cm, height=5cm,angle=0]{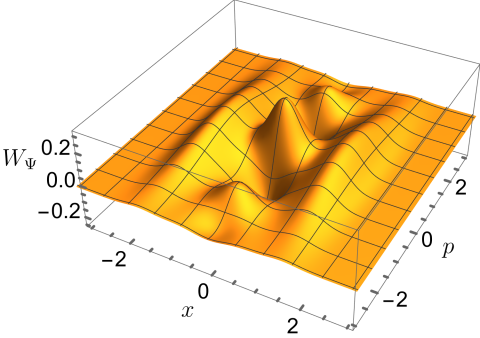} 
   &  \includegraphics[width=6cm, height=4.5cm,angle=0]{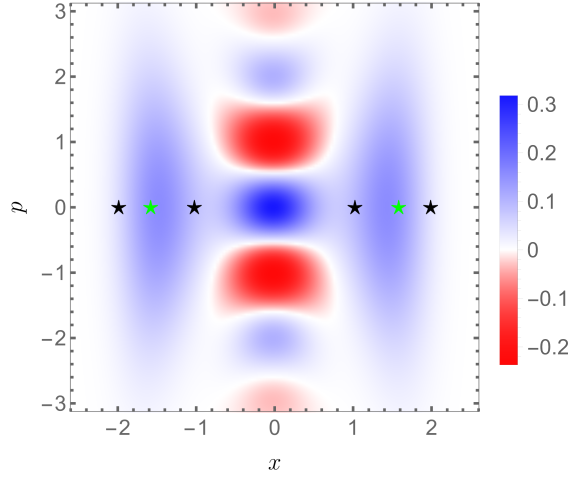} 
   \\ 
    \\ \hline
    \end{tabular}
    \caption{\label{Fig5}
    Wigner functions (left column) of the ground state of the BEC in a double well trap and its density (right column) for the case $\beta=0$ at $a=2$ (top row) and $5$ (bottom row), respectively. The black stars represent the classical turning points, and the green ones  the bottom of the degenerated wells.
    }
\end{figure*}

\begin{figure*}[t!]
    \begin{tabular}{ccc}
    \includegraphics[width=5cm, height=4cm,angle=0]{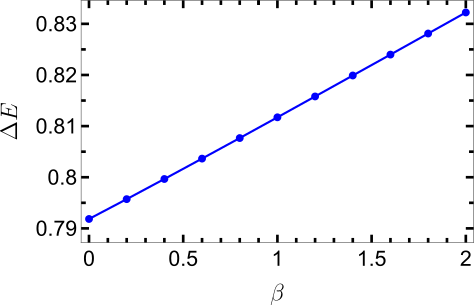} 
   &  \includegraphics[width=5cm, height=4cm,angle=0]{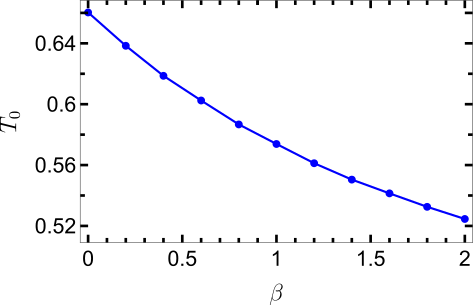} 
    \includegraphics[width=5cm, height=4cm,angle=0]{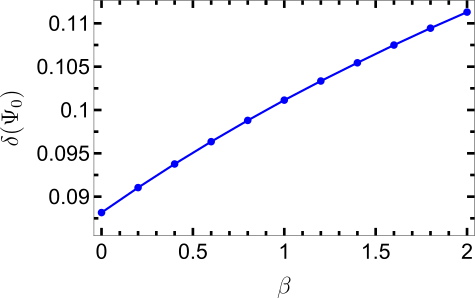} \\ 
        \includegraphics[width=5cm, height=4cm,angle=0]{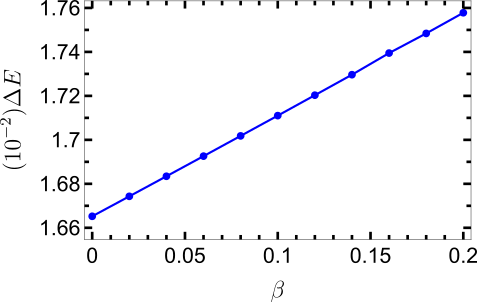} 
   &  \includegraphics[width=5cm, height=4cm,angle=0]{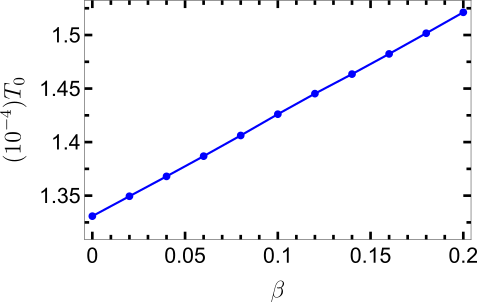} 
    \includegraphics[width=5cm, height=4cm,angle=0]{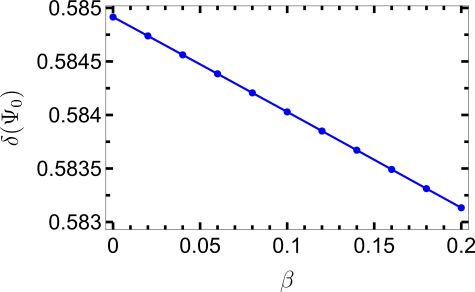} \\ 
    \hline
    \end{tabular}
    \caption{\label{Fig6} Volume of the negative part of the Wigner function $\delta(\Psi_0)$, energy splitting $\Delta E$ of the ground and first excited states, and transmission coefficient $T_0$ as a function of the parameter $\beta$. The rows correspond to $a=2$ (top) and $5$ (bottom), respectively.
    }
\end{figure*}

\begin{figure}[t!]
\centerline{\includegraphics[width=0.45\textwidth]{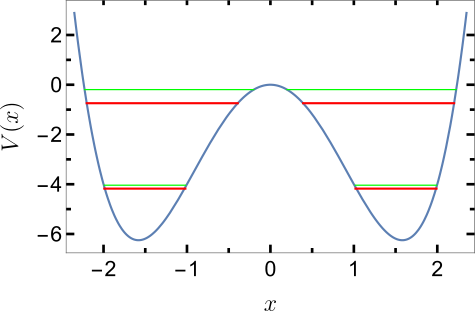}}
\caption{\label{Fig7} First four eigenenergies of the BEC condensate for $(a \, ,\beta )=  ( 5 , 0.1 )$. The eigenstates with even(odd) parity are denoted in red (green).}
\end{figure}

\begin{figure}[t!]
\centerline{\includegraphics[width=0.45\textwidth]{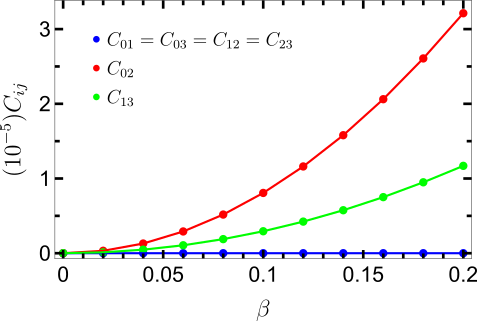}}
\caption{\label{Fig8} $\beta$-dependency of the overlaps among the four eigenfunctions with the lowest energies $C_{ij}=|\langle \Psi_i|\Psi_j \rangle |^2$ where $i=1,2,3,4$. The parameter of the double well is fixed at $a=5$.}
\end{figure}

The local maximum of the double well (\ref{Fig6}) at $x=0$ represents  a critical energy $E_c=0$ for the Schrödinger equation ($\beta=0$)~\cite{NADER2023129014} since the properties of the spectrum and the quantum dynamics change 
if the energy a certain state lies below or above the local maximum. In particular, when the energy is less than the local maximum of the potential, the central barrier represents a classically forbidden region that can be penetrated only via quantum tunneling.
It is well known that the energy of the ground state might be either positive or negative depending on the depth of the double well \cite{gonzalez2023parameter}. As the parameter $a$ increases, there is a particular value for which the energy of the ground state passes from positive to negative, i.e., the energy becomes less than the local maximum. This transition is accompanied by a qualitative change in the wave function which goes from one single peak, with maximum at $x=0$, to a double peak and being now the origin $x=0$ a minimum. The critical parameter $a_c$ can be estimated either by analysing the sign of the ground state energy or the sign of the second derivative of the wave function evaluated at $x=0$. By the numerical calculations plotted in Fig.~\ref{Fig1} of both quantities as functions of $a$, we conclude that both criteria match and the critical parameter is located at $a_c\sim1.7616$. By a direction calculation, we corroborate that the corresponding critical energy $E_c$ matches with the energy of the local maximum of the double well $E_c=0$. Since the critical energy corresponds to an unstable equilibrium point in the classical picture, with positive Lyapunov exponent~\cite{NADER2023129014}, it has some interesting implications on the quantum dynamics~\cite{Cary} which are summarized in Fig.~\ref{Fig2} (top-left plot). The consequences of an unstable point in the phase space are illustrated in Fig.~\ref{Fig2} (bottom plot) by the time evolution of a coherent state initially centered in the middle point of the separatrix corresponding to $(x_0 ,p_0 ) = (0,0)$. We observe that its density probability starts to spread over the whole separatrix. Additionaly, initial quantum states close to the separatrix shows an exponential growth of the Fidelity Out-of-time order Correlator (FOTOC) $F_{otoc}(t) =\sigma_x (t) + \sigma_p (t)$ which is equal to the addition of the variance of $x$ and $p$. We can observe in Fig.~\ref{Fig2} (top-right plot) that its slope matches with the positive Lyapunov exponent associated to the unstable point. The Density of States (DOS) will exhibit a singularity at the critical energy~\cite{NADER2023129014} which is the signature of an Excited State Quantum Phase Transition (ESQPT)~\cite{cejnar2021excited}. It is known that ESQPTs play an essential role in the quantum tunneling of initial states \cite{Nader2021PRA,Prado2023}.

We now consider the GP equation of a condensate gas (\ref{GP}) with finite atom-atom interaction $\beta>0$. Here, the critical parameter $a_c$ can only be estimated by the criterion of the second derivative because the \emph{effective} potential of the atoms contains the additional positive interaction $\beta |\Psi|^2$, and hence the critical energy $E_c$ may be different than its value in the Schr\"odinger scenario $E_c=0$. In Fig.~\ref{Fig3}, we show how the critical parameter is shifted for different values of $\beta$. One can associate a critical energy in the GP scenario by evaluating Eq.~\eqref{energy} at the critical parameter $a_c$ and its respective wavefunction. We plot in Fig.~\ref{Fig4} the behavior of the $a_c$ and $E_c$ as functions of $\beta$. As we expected, such transitions occur in general at critical energies greater than zero $E_c>0$.
\subsection{Nonclassicality and Tunneling}
Fig.~\ref{Fig5} shows the Wigner function~\eqref{Wigner} of the ground state of the  BEC in the double-well trap for different depths (controlled by the parameter $a$) in the case of noninteracting bosons ($\beta=0$). One can appreciate that the negative regions of the Wigner function lies along the coordinate axis. For small values of $a$ (see top row of Fig.~\ref{Fig5} for $a=2$), the Wigner function contains only one peak with shoulders centered at the local minima which eventually becomes in separated peaks if the double well is deep enough (bottom row for $a=5$). We now calculate the dependence of the nonclassicality of the ground state $\delta (\Psi_0)$~\eqref{Eqn.nonclass} with respect to $\beta$ (see Fig.~\ref{Fig6}). It is worth to mention that, even in the absence of boson-boson interactions $(\beta= 0)$, the nonlinear forces of the double-well potential produce negative regions of the Wigner function~\cite{HUDSON1974}. We observe in Fig.~\ref{Fig6} that the volume of the integrated negative part of the Wigner function changes as a function of the parameter $\beta$, i.e., the nonclassical behavior of the BEC is affected with the interaction between bosons. However, the monotonous deviation of $\delta (\Psi_0)$ with respect to $\beta$ depends on the parameter $a$.

The $\beta$-dependency of the nonclassicality reveals some information of the tunneling rate of the BEC~\cite{Yin2020}. To scrutinize this, we analyse the energy gap between the ground and the first excited states $\Delta E = E_1 - E_0 $, and the transmission coefficient in Fig.~\ref{Fig6} as a function of $\beta$ for the values of $a=2$ and 5. For the case of smaller well depth $a\sim2$, $\Delta E$ increases roughly linearly with respect to $\beta$ while the coefficient of transmission decrease. One can comprehend this by considering the enlargement of the effective barrier width encountered by the eigenstate, thereby impeding its ability to penetrate it. On the other hand, for $a\sim5$, while $\Delta E$ increases again roughly linearly, the transmission coefficient increase with respect to $\beta$ too. A potential explanation is that the width of the effective potential is barely modified at the bottom of the well, and then its penetration is enhanced with the interaction between bosons.  Contrary to the Schrödinger scenario $(\beta=0)$, the energy gap and the transmission coefficient are not proportional quantities for $\beta>0$.

For $a=5$, the double well is deep enough to contain four eigenstates below the critical. The energy levels are presented schematically in Fig.~\ref{Fig6}. It is observed that the first pair of energy states are quasidegenerated and close to the bottom of the double well. The energy gap increases for the second pair of energy levels since they are closer to the local maximum. A similar behavior was observed in the case of Gaussian wells~\cite{Schmelcher2008}. 
\subsection{Nonorthogonality of eigenstates}
Nonlinear quantum mechanics, such as the GP equation, might have nonorthogonal eigenstates~\cite{10.1063/1.531829}. This can be understood by the fact that the extra nonlinear terms in the corresponding nonlinear Schrödinger equation act as an extra effective potential. In the case of the GP equation~\eqref{GP}, the effective potential includes $\beta |\Psi|^2$. Consequently, the eigenstates come from different effective potentials. We plot in Fig.~\ref{Fig8} the overlap between the first eigenstates $C_{ij}=|\langle \Psi_i|\Psi_j \rangle |^2$ for $a=5$ as a function of $\beta$. We observe the unchanged orthogonality of states with different parity. On the contrary, states with the same parity become less orthogonal as $\beta$ increases. The nonorthogonality of the eigenstates is a signature of a nonlinear unitary evolution, which are been used in algorithms to distinguish nonorthogonal states, to solve the unstructured search problem~\cite{PhysRevA.93.022314}, and to devise nonunitary quantum gates such as feasible nonlinear Hadamard gates~\cite{PhysRevResearch.4.023071}.
\section{Conclusions and perspectives}
\label{Sec5.Conclusions}
In this work, we fully exploit the Finite Difference method to calculate the eigenstates and study the underlying physics of the 1D BECs confined in a double-well potential. The method is easily generalized for other confinement potentials. In particular, we have used it to study the deviation of the critical parameter $a_c$ and critical energy $E_c$ with respect to the strength of the interparticle interaction $\beta$. We conclude that the corresponding critical energy in general does not match with the local maximum of the double well. This shift may have some consequences on the dynamics since the local maximum of the effective barrier is higher than that of the double-well trap, and the corresponding separatrix of the classical phase-space corresponds to the shifted critical energy. Additionally, we scrutinize the tendency, as $\beta$ increases, of the nonclassicality, the energy gap between consecutive energy levels, the tunneling transmission coefficient through the energy barrier, and the nonorthogonality of the eigenstates. 

It is worth pointing out that for greater well depth $a$ and boson-boson interactions $\beta$, the eigenvalues converge to a particular value despite that the wavefunction not~\eqref{eq.conver}. Moreover, the wavefunction $\Psi_n^{(k)}$ jumps between the localized solutions over the two minima of the double-well potential in each iteration step of the selfconsistent method. An analog difficulty can be found in the Iterative Hartree—Fock Procedure which involves recalculation of the one‐electron density matrix
 ~\cite{koutecky1971convergence}. We pursue the implementation of the techniques to fix these issues or establish an upper bound for the non-linearity~\cite{Yang2009} in a forthcoming publication.
\section{Data Availability}
All codes, scripts, supplemental formulas and data needed to reproduce the results in this manuscript are available online  
{https://github.com/djuliannader/GrossPitaevskiiSCF}.
\section*{ACKNOWLEDGEMENTS}
ESE acknowledges support from the postdoctoral fellowship of the IPD-STEMA program of the University of Liège (Belgium).
\bibliographystyle{apsrev4-2}
\bibliography{BEC2023}
\end{document}